# Emission enhancement in dielectric nanocomposites

D. JALAS,[1,*] K. MARVIN SCHULZ,[1] A. YU. PETROV,[1,2] AND M. EICH[1,3]

*1 Hamburg University of Technology, Institute of Optical and Electronic Materials, Hamburg, Germany*
*2 ITMO University, St. Petersburg, Russia*
*3 Institute of Materials Research, Helmholtz-Zentrum Geesthacht, Geesthacht, Germany*
*\*dirk.jalas@tuhh.de*

**Abstract:** We consider emitting nanoparticles in dielectric nanocomposites with varying refractive index contrast and geometry. For that we develop a simple and universal method to calculate the emission enhancement in nanocomposites that employs solely the calculation of the effective refractive index and electric field distributions from three quasistatic calculations with orthogonal polarizations. The method is exemplified for dilute nanocomposites without electromagnetic interaction between emitting particles as well as for dense nanocomposites with strong particle interaction. We show that the radiative decay in dielectric nanocomposites is greatly affected by the shape and arrangement of its constituents and give guidelines for larger enhancement.

## References and links

1. Anger, P. Bharadwaj, and L. Novotny, "Enhancement and Quenching of Single-Molecule Fluorescence," Phys. Rev. Lett. **96**, 113002 (2006).
2. K. L. Tsakmakidis, R. W. Boyd, E. Yablonovitch, and X. Zhang, "Large spontaneous-emission enhancements in metallic nanostructures: towards LEDs faster than lasers [Invited]," Opt. Express **24**, 17916–17927 (2016).
3. E. M. Purcell, "Spontaneous emission probabilities at radio frequencies," Phys. Rev. **69**, 681 (1946).
4. S. Ek, P. Lunnemann, Y. Chen, E. Semenova, K. Yvind, and J. Mork, "Slow-light-enhanced gain in active photonic crystal waveguides," Nat. Commun. **5**, 5039 EP - (2014).
5. G. M. Akselrod, C. Argyropoulos, T. B. Hoang, C. Ciracì, C. Fang, J. Huang, D. R. Smith, and M. H. Mikkelsen, "Probing the mechanisms of large Purcell enhancement in plasmonic nanoantennas," Nat. Photonics **8**, 835–840 (2014).
6. T. Senden, F. T. Rabouw, and A. Meijerink, "Photonic Effects on the Radiative Decay Rate and Luminescence Quantum Yield of Doped Nanocrystals," ACS Nano **9**, 1801–1808 (2015).
7. K. Dolgaleva, R. W. Boyd, and P. W. Milonni, "Influence of local-field effects on the radiative lifetime of liquid suspensions of Nd:YAG nanoparticles," J. Opt. Soc. Am. B **24**, 516–521 (2007).
8. D. Toptygin, "Effects of the Solvent Refractive Index and Its Dispersion on the Radiative Decay Rate and Extinction Coefficient of a Fluorescent Solute," Journal of Fluorescence **13**, 201–219 (2003).
9. L. Rogobete, H. Schniepp, V. Sandoghdar, and C. Henkel, "Spontaneous emission in nanoscopic dielectric particles," Opt. Lett. **28**, 1736–1738 (2003).
10. A. H. Sihvola, *Electromagnetic mixing formulas and applications* (Iet, 1999).
11. A. E. Krasnok, A. P. Slobozhanyuk, C. R. Simovski, S. A. Tretyakov, A. N. Poddubny, A. E. Miroshnichenko, Y. S. Kivshar, and P. A. Belov, "An antenna model for the Purcell effect," Scientific Reports **5**, 12956 EP - (2015).
12. A. N. Poddubny, P. A. Belov, and Y. S. Kivshar, "Spontaneous radiation of a finite-size dipole emitter in hyperbolic media," Phys. Rev. A **84**, 23807 (2011).
13. P. Bharadwaj, B. Deutsch, and L. Novotny, "Optical Antennas," Adv. Opt. Photon. **1**, 438–483 (2009).
14. J. D. Jackson, *Classical electrodynamics*, 3rd ed (Wiley, 1999).
15. D. J. Griffiths, *Introduction to Electrodynamics* (Pearson Education, 2014).
16. Kumar, G. A., Chen, C. W., J. Ballato, and Riman, R. E., "Optical Characterization of Infrared Emitting Rare-Earth-Doped Fluoride Nanocrystals and Their Transparent Nanocomposites," Chem. Mater. **19**, 1523–1528 (2007).
17. Li, H. H., "Refractive index of alkaline earth halides and its wavelength and temperature derivatives," Journal of Physical and Chemical Reference Data **9**, 161–290 (1980).
18. M. Daimon and A. Masumura, "High-accuracy measurements of the refractive index and its temperature coefficient of calcium fluoride in a wide wavelength range from 138 to 2326 nm," Appl. Opt. **41**, 5275–5281 (2002).
19. T. Bellunato, M. Calvi, C. Matteuzzi, M. Musy, D. L. Perego, and B. Storaci, "Refractive index of silica aerogel: Uniformity and dispersion law," RICH 2007 **595**, 183–186 (2008).

20. A. Taflove, A. Oskooi, and S. G. Johnson, *Advances in FDTD computational electrodynamics: photonics and nanotechnology* (Artech house, 2013).
21. M. Pelton, J. Vukovic, G. S. Solomon, A. Scherer, and Y. Yamamoto, "Three-dimensionally confined modes in micropost microcavities: quality factors and Purcell factors," IEEE Journal of Quantum Electronics **38**, 170–177 (2002).
22. T. J. Kippenberg, Tchebotareva, A. L., J. Kalkman, A. Polman, and K. J. Vahala, "Purcell-Factor-Enhanced Scattering from Si Nanocrystals in an Optical Microcavity," Phys. Rev. Lett. **103**, 27406 (2009).
23. I. Iorsh, A. Poddubny, A. Orlov, P. Belov, and Y. S. Kivshar, "Spontaneous emission enhancement in metal–dielectric metamaterials," Physics Letters A **376**, 185–187 (2012).
24. A. A. Krokhin, P. Halevi, and J. Arriaga, "Long-wavelength limit (homogenization) for two-dimensional photonic crystals," Phys. Rev. B **65**, 115208 (2002).

## 1. Introduction

The radiative lifetime of the emitter influences the efficiency of a light source as there are always competing non-radiative processes present which dissipate energy [1]. The shorter the radiative lifetime is compared to other decay processes, the more likely the emitter's energy is transformed into light and not into heat. Besides the improvement of the efficiency of light sources, a fast radiative decay is also important for fast switching as the radiative decay rate sets the limit for the maximal modulation rate of a light source [2].

The radiative decay rate of an emitter, such as an erbium atom, is not only dependent on the emitter itself but can also be changed by modifying its environment. Strongly increased decay rates originate from a strong electromagnetic interaction between the luminescent medium and the optical field. Emission enhancement can be achieved in optical devices with local field enhancement such as resonators [3], slow light structures [4], or by focusing the optical fields into a small volume [5]. Of particular interest is the emission enhancement in nanocomposite media employing emitting nanoparticles and a host medium [6,7]. Typical considerations of this enhancement are limited to diluted spherical and elliptical inclusions [7,8] or flat slots [9] of an emitting medium in a high index host. There is no simple description available for dense nanocomposites of arbitrary geometry.

Commonly, the optical properties of dense dielectric nanocomposites are described by effective medium models [10]. These models fall short if one tries to calculate the optical emission as they can only describe the average electromagnetic fields. However, for emission processes local fields cannot be neglected. Here, we derive an equation for the average emission enhancement in dielectric nanocomposites based on the reciprocity relation and a quasistatic approximation. We discuss the enhancement for two shapes of luminescent particles in diluted composites as limiting cases and present enhancement calculations for two representative arrangements of the luminescent medium in dense composites using our approach. .

## 2. Calculating the emission of a dipole via reciprocity relation

Although the emission of a photon is a quantum mechanical process, we can accurately model the change of emission properties upon change of the environment by a classical dipole. The lifetime of a quantum emitter placed in vacuum divided by the lifetime inside the environment, $f_e = \frac{\tau_{vac}}{\tau_{enviroment}}$, is identical to the ratio of the power a continuously oscillating dipole emits inside the environment to the power emitted in vacuum, $f_e = \frac{P_{enviroment}}{P_{vac}}$ [11]. Thus, we will use the latter ratio as our metric for the emission enhancement.

The methods available to describe the emission of dipoles in structured media are usually based on Green's function approach [12]. For a composite, this approach requires complex numerical calculations for every point in the volume. In nanocomposite the structure parameters are usually much smaller than the optical wavelength. Here, we make use of this fact and utilize a simple quasistatic approximation in combination with the reciprocity relation. This consideration allows us to obtain the average emission enhancement for a distributed luminescent medium from three quasistatic calculations with orthogonal polarizations.

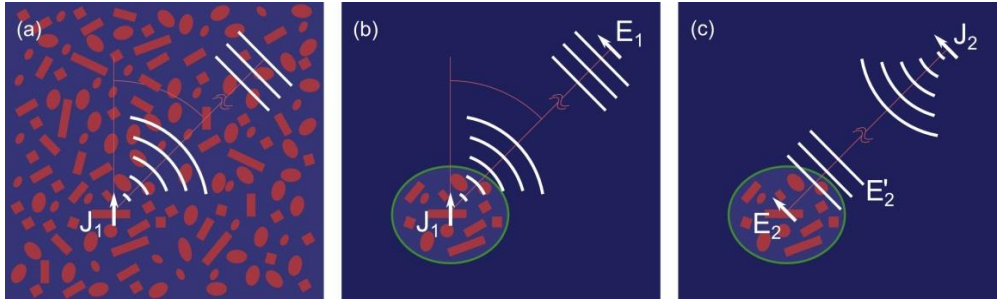

**Fig. 1.** (a) A dipole $J_1$ at $r = r_1$ emits into a structured medium. It produces a plane wave far away from the source at $r = r_2$. The emitting material is drawn in red. (b) The medium is artificially separated into two regions: nanostructured (inside the green perimeter) and homogeneous (outside). The plane wave has an amplitude $E_1$ at $r = r_2$. (c) Reversed case. A test dipole $J_2$ oriented along $E_1$ produces a field $E_2$ at the position of the original dipole and a field $E'_2$ at the green boundary.

The investigated structure is sketched in Fig. 1(a). It is a nanocomposite composed of an emitting material (drawn in red) inside a host material. We start by considering only a single emitter inside this nanocomposite. We model this emitter as an electric dipole. The power emitted by the dipole can be evaluated by integrating the intensity on the closed surface in the far field.

To calculate the far field emission of the source, we can simplify the problem by dividing the emission process into two parts: Initially, the dipole emits a wave and subsequently the wave propagates to our defined surface in the far field region. The wavelength of the propagating field is assumed to be much larger than the characteristic length of our nanocomposite and we can describe the wave propagation in the far field accurately with effective material parameters. We cannot do that for the dipole emission as this process is governed by the fields exactly at the position of the dipole. Therefore, we artificially divide the composite in two regions: In the outer region, we consider a macroscopically isotropic medium. In the inner region, we model the exact shape of the nanocomposite to obtain the electric field at the dipole position. The placement of the boundary was chosen such that it does not affect the fields at the emitter. The region outside has the same refractive index as the average refractive index of the nanocomposite, thus there will be no reflections at this boundary. Our goal is now to obtain the far fields emitted by the dipole such that we can quantify the emitted power.

We start by calculating the field, $E_1$, that the dipole produces far away from the source (see Fig. 1). An easy way to calculate the far field is to place a test dipole at the same positon and with the same orientation as the field we are interested in (Fig. 1(c)) [13]. This dipole will emit light as well and produce a field, $E_2$, at the initial dipole. We can connect the two cases in Fig. 1(b) and (c) with the help of the Rayleigh-Carson reciprocity theorem:

$$\int J_1 \cdot E_2 \, dV = \int E_1 \cdot J_2 \, dV. \tag{1}$$

With this relation in place we can use the second case to calculate the fields in the first case. For simplicity, we set $J_1 = |J_1| e_{J1} \delta(r - r_1)$ and $J_2 = |J_2| e_{J2} \delta(r - r_2)$ with $|J_1| = |J_2|$ and both current densities in phase. With this Eq. (1) becomes a simple scalar product

$$e_{J1} \cdot E_2(r_1) = E_1(r_2), \tag{2}$$

when $J_2$ is parallel to $E_1(r_2)$. We use the convention of vectors being represented by bold letters and scalar amplitudes being represented by light letters. Accordingly, if we know $E_2(r_1)$ we also know $E_1(r_2)$. In Fig. 1(c), $J_2$ creates a wave that has an amplitude of $E'_2 = \frac{1}{4\pi\epsilon_0} k_0^2 \frac{e^{jk(r_1-r_2)}}{|r_1-r_2|} \frac{|J_1|}{\omega}$, where $k_0 = \omega/c$, once it reaches the encircled region [14]. Far away from the source the curvature of the wave fronts can be neglected and the wave can be locally approximated by a plane wave with corresponding amplitude. Thus, to find the electric field at the position of the initial dipole we have to illuminate the encircled region with a plane wave and find the relation between $E_2(r_1)$ and $E'_2$. The ratio of the amplitudes of the two electric fields is the local field enhancement factor. It can be obtained by a full optical simulation. But nanostructured media have a characteristic length much smaller than the wavelength. Thus the plane wave excitation can be simplified to a quasistatic calculation with a constant bias field along the optical polarization. However the ratio between the outside and the inside field is obtained, we find that

$$E_1(r_2) = \frac{e_{J1} \cdot E_2(r_1)}{|E'_2|} \frac{1}{4\pi\epsilon_0} k_0^2 \frac{e^{jk(r_1-r_2)}}{|r_1-r_2|} \frac{|J_1|}{\omega}. \tag{3}$$

The intensity radiated in this direction is $I(r_2) = \frac{n_{eff}}{2Z_0} |E_1(r_2)|^2$, where $n_{eff}$ is the effective refractive index of the composite medium. Note that although we mentioned in the introduction that an effective refractive index is not sufficient to calculate the optical emission, we can use it here to model the wave propagation in the far field.

From the electric field in Eq. (3) we can determine the intensity at position $r_2$. We can calculate and sequentially integrate the intensity over a closed surface to obtain the emitted power, if we know $E_1$ at all positions on the surface. This would involve separate backward excitation for each direction. However, we show in the appendix that it is sufficient to illuminate the encircled region in Fig. 1 with three plane waves polarized in the three orthogonal orientations, $E'_{2,x}$, $E'_{2,y}$ and $E'_{2,z}$. Consequently, the power emitted by the dipole amounts to:

$$P_{rad,nc} = n_{eff} \left( \frac{|e_{J1} \cdot E_2^x|^2}{|E'_{2,x}|^2} + \frac{|e_{J1} \cdot E_2^y|^2}{|E'_{2,z}|^2} + \frac{|e_{J1} \cdot E_2^z|^2}{|E'_{2,z}|^2} \right) \frac{k_0^4}{12\pi Z_0 \varepsilon_0^2} \frac{|J_1|^2}{\omega^2}, \tag{5}$$

where $E_2^i$ is the local electrical field vector due to a illumination with $E'_{2,i}$. Note that $E_2^i$ is not necessarily aligned with $E'_{2,i}$ and thus we use $i$ as a superscript in $E_2^i$.

To obtain the emission enhancement factor, we need to divide this power by the power the emitter would emit into vacuum which is

$$P_{rad,vac} = \frac{k_0^4}{12\pi Z_0 \epsilon_0^2} \frac{|J_1|^2}{\omega^2}. \tag{6}$$

Thus, the emission enhancement factor is

$$f_{nc} = n_{eff} \left( \frac{|e_{J1} \cdot E_2^x|^2}{|E'_{2,x}|^2} + \frac{|e_{J1} \cdot E_2^y|^2}{|E'_{2,z}|^2} + \frac{|e_{J1} \cdot E_2^z|^2}{|E'_{2,z}|^2} \right). \tag{7}$$

Accordingly, the local field enhancement and the field's polarization have a strong effect on the radiative decay. The beauty of Eq. (7) is that we obtain the field enhancement factor for all emitter positions and orientations with only three quasistatic calculations.

Eq. (7) is useful in case we want to determine the emission of a dipole with a certain orientation. However, in many cases emitters are unpolarized. As we show in the appendix, the average emission enhancement of an unpolarized emitter still depends on the local field enhancement averaged for three orthogonally polarized incident fields $E'_{2,x}$, $E'_{2,y}$ and $E'_{2,z}$ and looks similar to the polarized case:

$$f_{unpol} = \frac{1}{3} n_{eff} \left( \frac{|E_2^x|^2}{|E'_{2,x}|^2} + \frac{|E_2^y|^2}{|E'_{2,y}|^2} + \frac{|E_2^z|^2}{|E'_{2,z}|^2} \right). \tag{8}$$

Just as for the polarized emitter case, $E_2^x$ is the local electrical field due to a biasing field in x-direction $E'_{2,x}$ and $E_2^x$ does not necessarily point in x-direction, as well. This equation allows calculating the emission enhancement for a complex composite using just three simulations with orthogonally polarized excitations.

Eqs. (7) and (8) describe a convenient way to calculate the emission enhancement of nanocomposites. In summary, the following two steps are needed: Firstly, one obtains the effective refractive index from Maxwell Garnett theory [10], a numerical simulation or any other suited method. Secondly, one chooses a representative volume around the emitter and calculates the electric field at the position of the emitter due to a static electric bias along the three principal directions. (If the volume features rotational symmetries, calculations for only one or two directions are sufficient.) The fields and the effective index are subsequently plugged into Eq. (7) in case of a polarized emitter or into Eq. (8) in case of an unpolarized emitter. In the following we give a few application examples for this method.

## 3. Diluted nanocomposites

As a first example, we calculate the emission enhancement of a diluted nanocomposite. The term *diluted* means two things in this example. Firstly, the emitting nanoparticles are so far away from each other such that they do not interact by their near fields. Secondly, the volume fraction of the nanoparticles is so small that the effective refractive index of the composite is equal to that of the host material.

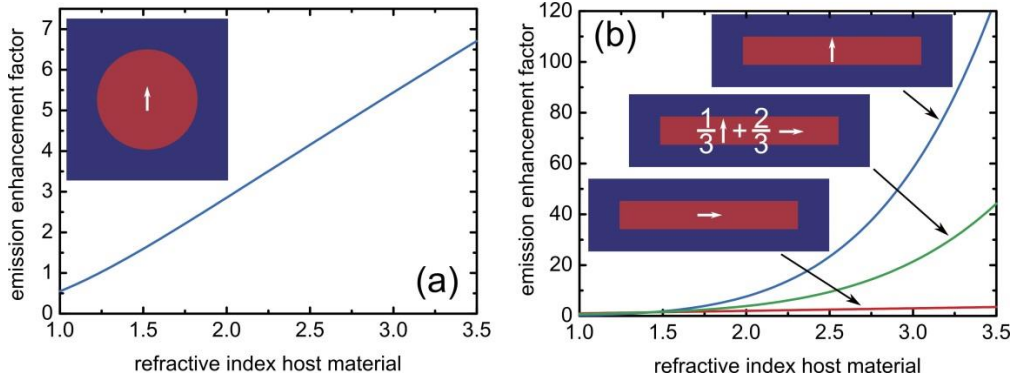

**Fig. 2.** (a) Emission enhancement factor vs host refractive index of an emitter in a spherical particle (Eq. (10)). (b) Emission enhancement factor vs host refractive index of an emitter in a platelet. Eq. (13) is the blue line, Eq. (14) the red and Eq. (15) the green. The luminescent material has refractive index 1.43 is colored red in the sketch.

To illustrate the importance of the nanoparticle shape on the emission we compare a sphere and a platelet with large aspect ratio. To calculate the emission enhancement factor of a dipole inside a sphere we use Eq. (7) for which we need to know the field enhancement factor. Since we consider a sphere which is much smaller than the wavelength, the field is practically in phase over the whole volume and the problem is equivalent to the case of a sphere biased with a static electric field. In this quasistatic case the field inside the sphere is related to the outside biasing field by [15],

$$E_2 = \frac{3n_{host}^2}{2n_{host}^2 + n_{emitter}^2} E'_2, \tag{9}$$

where $n_{host}$ is the refractive index of the host material and $n_{emitter}$ that of the particle. The field is constant inside the sphere. Inserting Eq. (9) back into Eq. (7) gives the emission enhancement factor

$$f_{sphere} = n_{host} \frac{|E_2|^2}{|E'_2|^2} = n_{host} \left(\frac{3n_{host}^2}{2n_{host}^2 + n_{emitter}^2}\right)^2, \tag{10}$$

where without loss of generality we chose $\boldsymbol{E}'_2$ to be parallel to $\boldsymbol{e}_{J1}$ such that the sum in Eq. (7) reduces to a single fraction. In literature, this factor inside the brackets of Eq. (10) is known as the local field correction factor from the real cavity model (see e.g. Refs. [7] and [8]). Note that the result is particle size independent as we are in the quasistatic regime and optical phase effects can be neglected. Because of the rotational symmetry of the sphere and the space independence of the field factor in Eq. (9), every dipole with arbitrary orientation and position inside the sphere will have the same field enhancement. Thus, Eq. (10) is not only the emission enhancement factor for a single dipole inside a nanosphere but also the averaged emission enhancement factor of a nanosphere made out of an emitting material.

Due to the reduced symmetry, the platelet consideration is slightly more complicated. We assume that the platelet has a large aspect ratio such that we can neglect fringe fields at the edges and model it as an infinitely extended slab. The field inside the platelet is then defined by the continuity condition at the platelet-host interfaces. For an electrical field perpendicular to the interfaces we have

$$\boldsymbol{E}_{2,\perp} = \frac{n_{host}^2}{n_{emitter}^2} \boldsymbol{E}'_{2,\perp} \tag{11}$$

and for fields parallel to the interfaces we have

$$\boldsymbol{E}_{2,\parallel} = \boldsymbol{E}'_{2,\parallel}. \tag{12}$$

Because the dipole sees only the fraction of the field that is parallel to it (Eq. (3)), a dipole oriented perpendicular to the interfaces has an emission enhancement factor of

$$f_{plate,\perp} = n_{host} \frac{|E_{2,\perp}|^2}{|E'_{2,\perp}|^2} = n_{host} \frac{n_{host}^4}{n_{emitter}^4} \tag{13}$$

On the other hand, a dipole parallel to the interfaces has no advantage to an unstructured material of refractive index $n_{host}$:

$$f_{plate,\parallel} = n_{host} \frac{|E_{2,\parallel}|^2}{|E'_{2,\parallel}|^2} = n_{host}. \tag{14}$$

As we neglect the fields at the edges, we obtain a position independent enhancement like we did for the sphere. In contrast to the sphere, the emission enhancement factor is polarization dependent. Thus, if we consider an unpolarized emitter we need to average over three orthogonal dipole orientations which yields for the platelet

$$f_{plate,unpol} = \frac{1}{3} n_{host} \frac{n_{host}^4}{n_{emitter}^4} + \frac{2}{3} n_{host}. \tag{15}$$

The platelet and the sphere show a strongly different dependence on the host material refractive index. The emission enhancement factor of the spherical nanoparticle scales linearly for a host material index much higher than the particle index, because the field enhancement converges to $(3/2)^2$. The emission enhancement factor scales linearly for emitters orientated parallel to the interface of the platelet. But the emission of perpendicular emitters scales with the fifth power of the host refractive index. We exemplify this shape dependence for erbium doped $CaF_2$ nanocrystals. These crystals emit at around 1.55µm optical wavelength [16] and have a refractive index of 1.43 [17,18]. We investigate how the emission enhancement factor changes if we place the particles inside host materials of varying refractive index. We vary the refractive index from 1 which would be a very porous material like an aerogel [19] to 3.5 which corresponds to silicon.

The resulting curves are shown in Fig. 2. For a spherical particle we see emission enhancement factors of up to 6.7 which is almost 5 times more than what can be achieved in $CaF_2$ alone. It is perhaps more practical to compare the emission in the nanostructured medium to the emission in a medium consisting purely out of the bulk host material instead of comparing to vacuum. This factor is $f_{e,bulk} = f_e/n_{host}$, i.e. in this case $f_{e,bulk} = f_e/1.43 = 4.7$. In stark contrast to the spherical particle the platelet can show an enhancement factor of $f_e = 125$ ($f_{e,bulk} = 87$). This value is reduced if one considers an unpolarized emitter as the parallel dipoles show only an enhancement scaling with $n_{host}$. Nonetheless, with an averaged enhancement of $f_e = 44$ ($f_{e,bulk} = 31$) the platelet emits 6.5 times stronger than the spherical particle. And the emission enhancement continuously grows proportionally to $n_{host}^5$.

## 4. Dense nanocomposites

We now turn to dense nanocomposites. In the case of a dense composite the volume fraction of the emitting particles is high enough such that they affect each other's radiation properties. Further, the emitting material cannot be neglected for the calculation of the effective refractive index of the nanocomposite. To describe emission in these structures we can use the same approach we used for the diluted nanocomposites but for the dense composites it is not justified to neglect fringe fields which are not aligned with the biasing electric field. We can use Eq. (8) and calculate the local fields for every position in the luminescent medium.

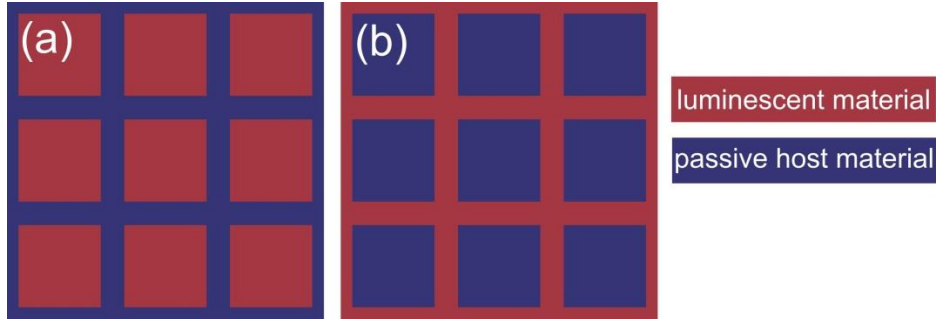

**Fig. 3.** We investigate two representative periodic 3D nanocomposites with 50/50 material split. (a) Emitting cubes inside a passive material. (b) Passive cubes inside an emitting material.

As an example for a dense composite, we investigate the two periodic composites shown in Fig. 3. The structure is in both cases a cubic crystal, whose unit cell consists of a cube surrounded by another material. In Fig. 3(a) the cubes are the emitting material and the surrounding medium is passive and in Fig. 3(b) we consider the reversed case. Again, the emitting material is erbium doped $CaF_2$ and the host material's refractive index is varied from 1 to 3.5. We chose a volume ratio of emitter material to host material of 50:50. To calculate the emission enhancement factor of the system we need to know the effective index of the system and the ratio of the local field to the biasing field. We obtain both via an eigenmode calculation with CST Microwave Studio. A detailed description of the simulation is given in the appendix. In contrast to the previous case, the field enhancement factor now spatially varys inside the emitting material and thus we calculate a volume averaged emission enhancement factor. Because the volume is much smaller than wavelength, the three static solutions yield all the possible optical states in the system. From which follows that our approach is, in this case, completely equivalent to calculating the emission enhancement factor via the local density of states [20].

One might expect that due the same amount of host and emitting material the average emission enhancement factor of the two arrangements in Fig. 3 is the same. As one can see in Fig. 4(a) this is not the case. The composite with emitting material inside the cubes scales linearly and thus resembles the behavior of the isolated sphere in Fig. 2(a). In contrast to that, the material with the emitter outside the cube shows a nonlinear dependence on the host refractive index. The reason for this behavior is similar to the emission enhancement of the platelet and can be seen in Fig. 4(b). If the emitting material is in between the cubes it forms platelet-like structures. The biasing field normal to them causes a field enhancement. The effect is not as strong as for the platelets because the fields are significantly reduced close to the edges of the cubes. Further, in contrast to the diluted composite, the effective refractive index of the dense composite is influenced by the emitting material. For a structure with $n_{host} = 3.5$, we obtain $n_{eff} = 2.2$, if the host medium is inside the cubes and $n_{eff} = 2.5$, if it is outside. The lower $n_{eff}$ in the first case is disadvantageous for the emission enhancement (see Eq. (8)), but from Fig. 4(a) we see that the local field enhancement exceeds the effect of the smaller $n_{eff}$ leading to a net enhancement.

The effect of the fringe fields can be significantly reduced if the volume fraction of the luminescent material is reduced. For a 10% volume fraction the emission enhancement for a host with refractive index of 3.5 can reach $f_e = 28$ ($f_{e,bulk} = 20$) if the host is concentrated in the cubes.

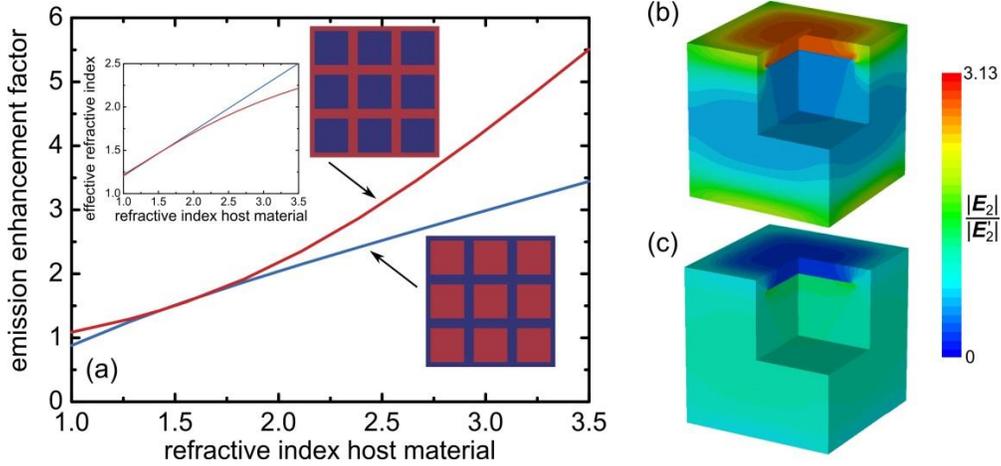

**Fig. 4.** (a) Emission enhancement factor vs refractive index of the passive material. For the blue line the emitting material is inside the cubes, for the red line its outside. The inset shows the effective refractive index of the two structures. (b) Electric field enhancement factor for a bias in vertical direction. Outside the cube is the emitting material with n=1.43 and inside is the passive material with n=3.5. (c) as (b) but with emitting material inside and passive material outside. Since a quasi-static approximation is validthe exact dimensions of the structure don't affect the result. For the fields in (b) and (c) an inner cube with 10nm long edges was considered.

## 5. Conclusion

In conclusion, we presented a method to calculate the emission enhancement in nanocomposites and applied this method to diluted and dense composites. While common effective medium models are not applicable and direct Green's function simulations are cumbersome, simple quasi static simulations can be used to calculate the average emission enhancement. Using the derived equations, we show that the emission can be greatly altered by the shape of the constituents and offers a route towards enhancing the performance of this material class.

Platelet based nanocomposites in high index materials like silicon can provide emission enhancement factors (compared to vacuum emission) in the order of 40, which is of the order of the enhancement factors achieved with high Q dielectric cavities [21,22]. Though higher emission enhancements can be obtained in small gap metal-dielectric resonators [5] and with metal-dielectric metamaterials [23], these structures as well as dielectric cavities require deliberated patterning on the scale of the wavelength or below and cannot easily be scaled up to yield large emitting volumes. The use of metals is also strongly connected to absorption losses. On contrary the presented novel approach can show large emission enhancement in bulk dielectric media even for luminescent material volume fraction up to 10%.

## Appendix 1: Derivation of average emission enhancement factor

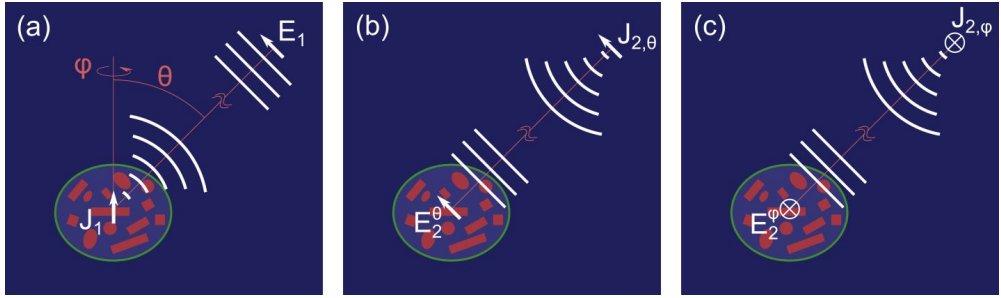

**Fig. 5.** (a) A dipole at $r = r_1$ emits into a structured medium. It produces the electric field far away from the source at $r = r_2$. The medium is artificially separated into two regions: nanostructured and homogeneous. (b) A test dipole oriented along $e_\theta$. (c) A test dipole oriented along $e_\varphi$. The emitting material is drawn in red.

Dipole 1 in Fig. 5(a) produces a far field in the direction $(\theta, \varphi)$ of $\boldsymbol{E}_1 = \boldsymbol{E}_{1,\theta} + \boldsymbol{E}_{1,\varphi}$. It produces no fields in the radial direction as only transverse waves propagate in the farfield.

We can calculate the fields in the two polarizations separately by placing test dipoles with respective orientation as indicated in Fig. 5(b) and (c). Like we did in the main text, we can relate the fields by

$$\boldsymbol{e}_{J1} \cdot \boldsymbol{E}_2^\theta(\boldsymbol{r}_1) = \boldsymbol{e}_\theta \cdot \boldsymbol{E}_1(\boldsymbol{r}_2) = E_{1,\theta}(\boldsymbol{r}_2), \qquad (16)$$

$$\boldsymbol{e}_{J1} \cdot \boldsymbol{E}_2^\varphi(\boldsymbol{r}_1) = \boldsymbol{e}_\varphi \cdot \boldsymbol{E}_1(\boldsymbol{r}_2) = E_{1,\varphi}(\boldsymbol{r}_2). \qquad (17)$$

We assumed that $|J_1| = |J_{2,\theta}| = |J_{2,\varphi}|$ and that all currents are in phase. We can again calculate the far fields $\boldsymbol{E}'_{2,\theta}$ and $\boldsymbol{E}'_{2,\varphi}$ and obtain $\boldsymbol{E}_2^\theta(\boldsymbol{r}_1)$ and $\boldsymbol{E}_2^\varphi(\boldsymbol{r}_1)$ from a quasistatic consideration. This approach would require a separate calculation for each direction $(\theta, \varphi)$. We can reduce the number of necessary calculations by expanding the fields at $\boldsymbol{r}_1$ in three excitations with orthogonal polarizations:

$$\boldsymbol{E}_2^\theta(\boldsymbol{r}_1) = \left( \cos(\theta)\cos(\varphi) \frac{\boldsymbol{E}_2^x}{|\boldsymbol{E}'_{2,x}|} + \cos(\theta)\sin(\varphi) \frac{\boldsymbol{E}_2^y}{|\boldsymbol{E}'_{2,y}|} - \sin(\theta) \frac{\boldsymbol{E}_2^z}{|\boldsymbol{E}'_{2,z}|} \right) \frac{1}{4\pi\epsilon_0} \left(\frac{\omega}{c_0}\right)^2 \frac{e^{jk(r_1-r_2)}}{|r_1-r_2|} \frac{|J_1|}{\omega}, \qquad (18)$$

$$\boldsymbol{E}_2^\varphi(\boldsymbol{r}_1) = \left( -\sin(\phi) \frac{\boldsymbol{E}_2^x}{|\boldsymbol{E}'_{2,x}|} + \cos(\phi) \frac{\boldsymbol{E}_2^y}{|\boldsymbol{E}'_{2,y}|} \right) \frac{1}{4\pi\epsilon_0} \left(\frac{\omega}{c_0}\right)^2 \frac{e^{jk(r_1-r_2)}}{|r_1-r_2|} \frac{|J_1|}{\omega}. \qquad (19)$$

Here, we utilize the superposition principle i.e. that in the quasistatic regime we can decompose the fields $\mathbf{E}_2^\theta(\mathbf{r}_1)$ and $\mathbf{E}_2^\varphi(\mathbf{r}_1)$ into three fields each caused by a biasing field in direction of one of the Cartesian axes. Note that $\mathbf{E}_2^x$, $\mathbf{E}_2^y$ and $\mathbf{E}_2^z$ represent local fields excited by three orthogonal polarizations, and thus they do not have to be orthogonal to each other. We can obtain $\mathbf{E}_{1,\theta}(\mathbf{r}_2)$ and $\mathbf{E}_{1,\varphi}(\mathbf{r}_2)$ by inserting Eqs. (18) and (19) into (16) and (17).

$$\mathbf{E}_{1,\theta}(\mathbf{r}_2) = \left(\cos(\theta)\cos(\varphi)\frac{e_{J1}\cdot\mathbf{E}_2^x}{|E'_{2,x}|} + \cos(\theta)\sin(\varphi)\frac{e_{J1}\cdot\mathbf{E}_2^y}{|E'_{2,y}|} - \sin(\theta)\frac{e_{J1}\cdot\mathbf{E}_2^z}{|E'_{2,z}|}\right) \frac{1}{4\pi\epsilon_0}\left(\frac{\omega}{c_0}\right)^2 \frac{e^{jk(r_1-r_2)}}{|r_1-r_2|}\frac{|J_1|}{\omega}, \quad (20)$$

$$\mathbf{E}_{1,\varphi}(\mathbf{r}_2) = \left(-\sin(\phi)\frac{e_{J1}\cdot\mathbf{E}_2^x}{|E'_{2,x}|} + \cos(\phi)\frac{e_{J1}\cdot\mathbf{E}_2^y}{|E'_{2,y}|}\right) \frac{1}{4\pi\epsilon_0}\left(\frac{\omega}{c_0}\right)^2 \frac{e^{jk(r_1-r_2)}}{|r_1-r_2|}\frac{|J_1|}{\omega}. \quad (21)$$

Because $\mathbf{E}_{1,\theta}(\mathbf{r}_2)$ and $\mathbf{E}_{1,\varphi}(\mathbf{r}_2)$ are orthogonal we can write the intensity as

$$I(\theta,\varphi) = \frac{n_{eff}}{2Z_0}\left(|\mathbf{E}_{1,\theta}|^2 + |\mathbf{E}_{1,\varphi}|^2\right). \quad (22)$$

To find the total radiated power we need to integrate the intensity over the whole solid angle

$$P_{dipole} = \int_0^{2\pi}\int_0^\pi \frac{n}{2Z_0}\left(|\mathbf{E}_{1,\theta}|^2 + |\mathbf{E}_{1,\varphi}|^2\right) r_1 \sin(\theta)\,\mathrm{d}\theta\mathrm{d}\varphi. \quad (23)$$

This yields

$$P_{dipole} = \frac{n_{eff}}{2Z_0}\frac{8\pi}{3}\left(\frac{1}{4\pi\epsilon_0}\left(\frac{\omega}{c_0}\right)^2\frac{|J_1|}{\omega}\right)^2\left(\frac{|e_{J1}\cdot\mathbf{E}_2^x|^2}{|E'_{2,x}|^2} + \frac{|e_{J1}\cdot\mathbf{E}_2^y|^2}{|E'_{2,z}|^2} + \frac{|e_{J1}\cdot\mathbf{E}_2^z|^2}{|E'_{2,z}|^2}\right). \quad (24)$$

With a vacuum space emission of $P_{rad,vac} = \frac{k_0^4}{12\pi Z_0 \varepsilon_0^2}\frac{|J_1|^2}{\omega^2}$ we obtain the emission enhancement factor of a single dipole of

$$f_{dipole} = n_{eff}\left(\frac{|e_{J1}\cdot\mathbf{E}_2^x|^2}{|E'_{2,x}|^2} + \frac{|e_{J1}\cdot\mathbf{E}_2^y|^2}{|E'_{2,z}|^2} + \frac{|e_{J1}\cdot\mathbf{E}_2^z|^2}{|E'_{2,z}|^2}\right). \quad (25)$$

The emission enhancement factor of an unpolarized emitter can be obtained by averaging the enhancement for three orthogonal emitters, $f_{unpol} = \frac{1}{3}(f_x + f_y + f_x)$. With $\sum_{i=x,y,z}|e_i\cdot\mathbf{E}|^2 = |\mathbf{E}|^2$ we obtain

$$f_{unpol} = \frac{1}{3}n_{eff}\left(\frac{|E_2^x|^2}{|E'_{2,x}|^2} + \frac{|E_2^y|^2}{|E'_{2,z}|^2} + \frac{|E_2^z|^2}{|E'_{2,z}|^2}\right). \quad (26)$$

### Appendix 2: Calculating the emission enhancement factor with an eigenmode solver

We can find the emission enhancement factor with a numerical eigenmode calculation. For this purpose we simulate a unit cell of our periodic structure with boundary conditions as indicated in Fig. 6. Since we are in the quasistatic regime and the mode is practically in phase over the whole unit cell, we set the phase difference $\Delta\phi$ to a small number, e.g. 0.1°. Slight changes of this value will proportionally change the mode frequency which is of no relevance since in the quasistatic regime the modal fields are frequency independent. We use a small non zero phase to obtain the local field distribution and the $n_{eff}$ from the same simulation.

We use the eigenmode solver of CST Microwave Studio. The solver returns the modes frequency and its fields normalized to carry 1J of electromagnetic energy in the simulation volume. To calculate the emission enhancement factor we need to obtain the effective index and the ratio between local field and biasing field.

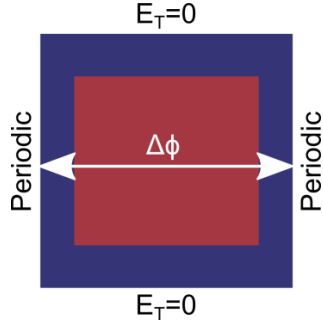

**Fig. 6.** Eigenmode solver boundary conditions: In the horizontal direction periodic boundary conditions with a phase shift of $\Delta\phi$ are used. In the vertical direction the tangential electric field is set to zero. In the remaining orthogonal direction, the tangential magnetic field is zero. These boundary conditions yield a plane wave traveling in horizontal direction with electric field polarized in vertical direction. To obtain the solution for another electric field polarization the boundary conditions need to be switched accordingly. Each volume edge has a length of a.

We can calculate the effective index from the k-vector:

$$k = \frac{\omega}{c_0} n_{eff} = \frac{\Delta\phi}{a}, \tag{162}$$

where $a$ is the volume length, $c_0$ is the vacuum speed of light and $\omega$ is the modes radial frequency. The biasing field is the average electric field of the wave traveling through the volume. We can calculate it from the power the mode carries:

$$P_{mode} = \frac{n_{eff}}{2Z_0} |\boldsymbol{E}_{avg}|^2. \tag{163}$$

The power can be either obtained by integrating the Poynting flux in or out of the volume or we can use the fact that the mode is normalized to 1J:

$$P_{mode} = \frac{v_g}{a} W_E \approx \frac{\omega}{c_0} n_{eff} \frac{1}{a} W_E, \tag{164}$$

where $v_g$ is the group velocity equal to phase velocity in the effective medium. This simplification can be made for periodic structures with periods much smaller than the wavelength if material dispersion can be neglected [24].

**Acknowledgments**

The authors want to acknowledge the funding provided by the German Federal Ministry of Education and Research (BMBF) for the grant 01GR0468. This work is a part of the 382-PiGnano project of ERA.Net RUS Plus 2013-2018 initiative under Consortium Agreement with Swiss Federal Laboratories for Materials Science and Technology (Switzerland) and Hamburg University of Technology (Germany). The publication was funded by the Deutsche Forschungsgemeinschaft (project no. 392323616) and the Hamburg University of Technology in the funding program Open Access Publishing. The authors acknowledge the support of CST, Darmstadt, Germany with their Microwave Studio software.